# 2-Sat Sub-Clauses and the Hypernodal Structure of the 3-Sat Problem

Douglas Powell

Abstract

Like simpler graphs, nested (hypernodal) graphs consist of two components: a set of nodes and a set of edges, where each edge connects a pair of nodes. In the hypernodal graph model, however, a node may contain other graphs [1]; so that a node may be contained in a graph that it contains. The inherently recursive structure of the hypernodal graph model aptly characterizes both the structure and dynamic of the 3-sat problem, a broadly applicable, though intractable, computer science problem. In this paper I first discuss the structure of the 3-sat problem, analyzing the relation of 3-sat to 2-sat, a related, though tractable problem. I then discuss sub-clauses and sub-clause thresholds and the transformation of sub-clauses into implication graphs, demonstrating how combinations of implication graphs are equivalent to hypernodal graphs. I conclude with a brief discussion of the use of hypernodal graphs to model the 3-sat problem, illustrating how hypernodal graphs model both the conditions for satisfiability and the process by which particular 3-sat assignments either succeed or fail.

## 1. Introduction and Terminology

In its standard form (*Conjunctive Normal Form, or CNF*), the *3*-sat problem consists of a formula $F$ which is a conjunction of a set of *m clauses $\{C_0, C_1, ..., C_{m-1}\}$*, where each clause is the disjunction of a set of three *literals* and each literal is the *true* or *false* instantiation of a member of the set of *n* Boolean *variables, $\{X_0, X_1, ..., X_{n-1}\}$*. The set of *n* true literals is represented by $\{x_0, x_1, ..., x_{n-1}\}$ and the set of *n* false literals is represented by $\{-x_0, -x_1, ..., -x_{n-1}\}$. The symbol $L$ refers to the set of true and false literals. The '-' symbol represents negation, so that, for example, $-(-x0) = x0$ and $-(x0) = -x0$. The '^' symbol represents conjunction and the 'v' symbol represents disjunction. Thus, a 3-sat formula has the form, $(C_0 \wedge C_1 \wedge ... \wedge C_{m-1})$, where each clause takes the form $(l_1 \vee l_2 \vee l_3)$ and $l_1$, $l_2$, and $l_3$ are true or false literals. Because contradictory pairs of literals are tautological and duplicate literals are redundant, the literals in a given clause are both distinct and non-contradictory. The 2-sat problem is defined similarly, though in 2-sat problems there are only two literals per clause.

Assignments

An assignment for a particular 3-sat problem consists of a set of literals drawn from the sets of true and false literals for that problem. A *consistent* assignment contains either $-x_i$ or $x_i$, but not both, for $i \in \{0...n-1\}$, whereas an *inconsistent* assignment contains both $-x_i$ and $x_i$ for some $i \in \{0...n-1\}$. Formally, an assignment $A$ is consistent if $\forall a \in A( -a \notin A )$ but *inconsistent* if $\exists a \in A( -a \in A )$. A *partial assignment* satisfies some subset of clauses in a 3-sat formula. Unless otherwise noted, in this paper the term *assignment* refers to a consistent (non-partial) assignment.

Satisfaction

A literal *satisfies* the clauses in which it appears. For a given 3-sat formula, each literal appears in and satisfies some subset of the clauses of the formula, whereas its negation satisfies a distinct subset of clauses. If $c$ is a clause in $F$, $L$ the complete set of literals such that $L = \{x_0...x_{n-1}, -x_0...-x_{n-1}\}$, and $a$ an arbitrary literal such that a $\in$ L, then the set of clauses satisfied by $a$ is given by the formula *satisfied(a)* = $\forall c \in F$ (a$\in$c). Likewise, *satisfied(-a)* = $\forall c \in F$ (-a$\in$c).

A clause is *satisfied* by an assignment (or partial assignment) when one or more of its three constituent literals is contained in that assignment. Thus, an assignment $A$ satisfies a clause $c$ iff $(c \wedge A) \neq \varnothing$. For example, an assignment containing -x0 satisfies the clause (-x0 v -x3 v -x4) but not the clause (x0 v -x3 v -x4).

A 3-sat problem is *satisfiable* if there is a consistent assignment to some set of the $n$ variables such that each clause is satisfied; conversely, the problem is *unsatisfiable* if no such assignment exists. Formally, an assignment $A$ satisfies a formula $F$ with clauses $C$ if $\forall c \in C( (c \wedge A) \neq \varnothing ) \wedge \forall a \in A( -a \notin A )$.

3-sat Algorithms, Time Complexity and 2-sat

A *complete* algorithm for the 3-sat problem either finds an assignment for the problem instance or determines that the problem is unsatisfiable (see Gu [2] for a comprehensive study of existing *sat* algorithms). In the worst case, the performance of existing complete 3-sat algorithms degrades to exponential time, taking. $2^n$ steps to solve the problem, unfeasible for even moderately large $n$ [3]. The performance of existing 2-sat algorithms is far better: there are complete 2-sat algorithms that solve the 2-sat problem in linear time–trivial for even large $n$ [4].

In general, the difficulty of a set of 3-sat problems is defined by the ratio of clauses to variables $r=m/n$ [5]. The problems used to investigate the concepts of sub-clauses and hypernodality were all generated with r=4.25, the region where roughly half the 3-sat problems are satisfiable and half are unsatisfiable [6].

## 2. Satisfied Clause Sets and 2-Sat Sub-Clauses

Consider the clause $c$ = (-x1 v -x4 v x5). Assignment to the negation of any of the three literals in the clause reduces the clause to a two-literal sub-clause. For example, an assignment of x1 reduces the clause $c$ to (-x4 v x5), whereas an assignment of x4 reduces $c$ to (-x1 v x5) and an assignment of -x5 reduces $c$ to (-x1 v -x4). Since each 3-sat clause contains three literals, each clause contains three potential 2-sat sub-clauses, derived by negating, one at a time, each of the three literals in the clause (fig.1). Thus, if *-a* is used in an assignment, each three-literal clause in *satisfied(a)* is reduced to a two-literal *sub-clause*, since $-a \wedge (a\ v\ l_2\ v\ l_3) = (l_2\ v\ l_3)$.

| Clause | ^ | Negated Literal | => | Sub-Clause |
|---|---|---|---|---|
| $c = (l_1 \vee l_2 \vee l_3)$ | | $-l_1$ | | $S_0 = \{l_2 \vee l_3\}$ |
| | | $-l_2$ | | $S_1 = \{l_1 \vee l_3\}$ |
| | | $-l_3$ | | $S_2 = \{l_1 \vee l_2\}$ |

**Figure 1**. Negated literals and sub-clauses.

The Sub-Clause Space, S

A 3-sat problem in $n$ variables contains $2n$ literals. Since a literal may not appear in a sub-clause with either itself or its negation, each literal may be combined with $2(n-1)$ literals, yielding a total of $2n(n-1)$ potential sub-clauses. If $r$ is the ratio of clauses to variables, then a given 3-sat problem $F$ contains $(r*n)$ clauses and $3r*n$ sub-clauses (since each clause contains 3 sub-clauses), assuming that there are no duplicate sub-clauses in $F$. Thus the ratio of actual to possible sub-clauses $= \frac{3r}{2(n-1)}$, so the probability that any two literals produce the same sub-clause diminishes with increasing $n$. Given a 3-sat formula $F$, the sub-clause space $S$ denotes the set of sub-clauses in $F$, equal to the union of the sub-clauses contained in the clauses of $F$.

The Relation of Literals to Sub-Clauses

Let $S$ denote the set of sub-clauses in a given 3-sat formula $F$. Each sub-clause in $S$ is activated by one or more literals, solved by either of its two member literals and converted into a unit clause (a clause containing one literal) by the negation of either of its two member literals. Similarly, each literal in a particular 3-sat problem activates a subset of the sub-clauses in $S$, solves a subset of those sub-clauses and converts a subset of those sub-clauses into unit clauses. Assuming that $L = \{x0...xn-1, -x0...-xn-1\}$, $a$ is an arbitrary literal such that $a \in L$, *satisfied(a)* is the set of clauses satisfied by $a$, $c$ is a clause in $F$, and $Q$ is an arbitrary set of sub-clauses, the sets described in Fig. 2 show the relation of literals to their related sub-clauses:

> The term **subclauses(a)** denotes the set of two-literal clauses, interpreted as a partial CNF 2-sat formula, created if $a$ is used in an assignment. Subclauses(a) is activated when $a$ is assigned and is created by removing -a from each member clause in satisfied(a). Formally, subclauses(a) = $\forall s \in$ satisfied(-a) (s $\wedge$ a). Likewise, given a set of literals $P$, *subclauses(P)*, and a literal $a \in P$, *subclauses(P)* is the union of *subclauses*(a).

> The term **subsat(a)** denotes the set of sub-clauses satisfied by $a$. Note that $a$ only satisfies sub-clauses that are activated. Formally, *subsat(a)* = $\forall s \in S (a \in s)$.

> The term **unitclauses(a)** refers to the set of literals derived by removing $a$ from *subsat(-a)*. Formally, *unitclauses(a)* = $\forall s \in$ *subsat(-a)* (s $\wedge$ a).

The term *creators(Q)* refers to the set of literals that create (or activate) the set of sub-clauses in Q. Formally, *creators(Q)* = ∀a ∈ L ( (subclauses(a) ∧ Q) ≠ ∅) .

The term *parent(s)* refers to the set of clauses from which the sub-clause *s* derives. Likewise, if *S* is a set of sub-clauses, then *parent(S)* is the union of *parent(s)* for s ∈ S.

**Figure 2.** Definitions relevant to sub-clauses.

The *satisfied()*, *subclauses()*, *subsat()* and *unitclauses()* sets are interdependent in that an assignment to a literal creates a set of sub-clauses, which appear in the *subsat()* sets of other literals, a subset of which much be assigned to satisfy the sub-clauses in *subclauses()*, thereby creating unit clauses, which force the use of other literals, which, in turn, create more sub-clauses. As an algorithm uses literals in an assignment, this cycle continues, until either a satisfying assignment is found, and all activated sub-clauses are solved, or a contradiction is created, and some sub-clauses are left unsolved. (In *dataflow* terms, each literal *produces* a set of sub-clauses that must then be *consumed,* via satisfaction, by other literals.) The sub-clause interaction matrix (Fig. 2b) summarizes the connectivity of literals and sub-clauses, showing how literals create, solve and convert sub-clauses to unit variables. For example, the literal *-x0* creates the sub-clause *s0*, which, when combined with *-x2,* creates the unit clause containing *-x3*.

|     | -x0 | x0  | -x1 | x1  | -x2 | x2  | -x3 | x3  |
| --- | --- | --- | --- | --- | --- | --- | --- | --- |
| s0  | c   |     | c   |     | -x3 | s   | s   | x2  |
| s1  | -x3 | s   |     | c   | c   |     | s   | x0  |
| s2  | x2  | s   |     | c   | x0  | s   | c   | c   |
| s3  |     | c   |     |     | s   | x3  | -x2 | s   |
| s4  | s   | x3  | c   | c   | c   | c   | -x0 | s   |
| s5  | s   | -x2 |     |     | s   | -x0 | c   |     |
| s6  | c   | c   | c   | c   | x3  | s   | x2  | s   |
| s7  | s   | x2  |     |     | -x0 | s   | c   |     |
| s8  |     | c   | x3  | s   | c   |     | x1  | s   |
| s9  |     |     | x2  | s   | x1  | s   | c   | c   |
| s10 |     | c   | s   | x3  | c   |     | -x1 | s   |
| s11 | s   | -x1 | s   | -x0 |     |     | c   | c   |
| s12 |     |     | c   | c   | s   | -x3 | s   | -x2 |
| s13 |     |     | -x3 | s   | c   | c   | s   | x1  |
| s14 |     |     | -x2 | s   | s   | x1  |     | c   |
| s15 | s   | x1  | -x0 | s   |     |     | c   |     |
| s16 | c   | c   | s   | -x3 |     | c   | s   | -x1 |
| s17 | s   | -x3 |     | c   |     |     | s   | -x0 |
| s18 |     |     | s   | -x2 | s   | -x1 |     | c   |
| s19 | c   |     | s   | x2  | -x1 | s   | c   |     |
| s20 | -x1 | s   | s   | x0  | c   |     |     | c   |
| s21 | x3  | s   |     |     | c   |     | x0  | s   |

**Figure 2b.** Interaction Matrix for Literals and Sub-Clauses. Key: c = sub-clause created by literal; s = sub-clause solved by literal; a literal in a cell is a unit variable created by the combination of the corresponding row sub-clause and column literal (problem=k3n4Seq207p0).

As an example of the interaction of clauses, literals and sub-clauses, suppose that $F$ contains the clauses $C_0 = \{-x1 \vee -x4 \vee x5\}$, $C_1 = \{-x3 \vee -x4 \vee x8\}$ and $C_2 = \{-x1 \vee -x3 \vee -x4\}$, which, through assignment of x4, become the sub-clauses $s_0 = \{-x1 \vee x5\}$, $s_1 = \{-x3 \vee x8\}$ and $s_2 = \{-x1 \vee -x3\}$. Then, since

    *satisfied(-x4)* $\quad = \{C_0, C_1, C_2\}$,
    *subclauses(x4)* $\quad = \{ \{C_0 \wedge x4\}, \{C_1 \wedge x4\}, \{C_2 \wedge x4\}\} = \{ \{-x1 \vee x5\}, \{-x3 \vee x8\},$
                                              $\{-x1 \vee -x3\} \}$
                                 $= \{s_0, s_1, s_2\}$,
    *subsat(-x3)* $\quad = \{s_1, s_2\}$, and
    *unitclauses(x3)* $\quad = \{\{s_1 \wedge x3\}, \{s_2 \wedge x3\}\} = \{\{x8\}, \{-x1\}\} \approx \{x8, -x1\}$.

In summary, an assignment of x4 activates sub-clauses $\{s_0, s_1, s_2\}$, which, in combination with an assignment to x3 yield unit clauses containing -x1 and x8, which must thus be part of the assignment.

### 3. Sub-clause Thresholds and the use of Sub-clauses

<u>Sub-Clause Thresholds</u>

The set of sub-clauses for a formula is given by $\forall a \in L( \textit{subclauses}(a) )$. Given the set of variables $V = \{X_0, X_1, ..., X_{n-1}\}$, an assignment contains either a true or false literal for each variable. Assuming that no two literals create the same sub-clause, the minimum number of sub-clauses in an assignment for the formula is given by

    minimum_threshold = $| \forall x \in V( \min(|\textit{subclauses}(x)|, |\textit{subclauses}(-x)|) |$,

and the maximum number of sub-clauses in an assignment for the formula is given by

    maximum_threshold = $| \forall x \in V( \max(|\textit{subclauses}(x)|, |\textit{subclauses}(-x)|) |$

If the sub-clause count for an assignment $A$ is given by

    subclause_count = $| \forall a \in A( \textit{subclauses}(a) ) |$,

then, for any satisfying assignment,

    minimum_threshold <= subclause_count <= maximum_threshold.

Given a 3-sat formula $F$, any satisfying assignment $A$ must satisfy at least as many sub-clauses as are in the minimum threshold, and at most as many sub-clauses as are in the maximum threshold. The rate at which a satisfying assignment must ultimately satisfy sub-clauses is given by the formula: $| \forall a \in A( \textit{subclauses}(a) ) | \div | A |$, which is equal to $| \forall a \in A( \textit{subclauses}(a) ) | \div n$. Finally, if an assignment $A$ satisfies a 3-sat formula, then: $(\forall a \in A( \textit{subclauses}(a) )) = (\forall a \in A( \textit{subsat}(a) ))$.

Four Types of Sub-Clause-Derived Assignments

By comparing the number of sub-clauses created and solved by the true and false literals of a particular variable, it is trivial to derive the following four types of assignments: 1) minCreate, whose literals have the minimum number of created sub-clauses, 2) minCreateMaxSolve, whose literals are the lesser of ( | solved sub clauses | - | created sub clauses | ), 3) maxSolve, whose literals solve the maximum number of sub-clauses, and 4) maxCreate, whose literals create the maximum number of sub-clauses. In measures of assignment correctness (% clauses satisfied, % created but unsatisfiable sub-clauses, etc.), assignment types 1-3 are roughly equivalent to assignments generated using a simple greedy algorithm [7] which selects variables based on per-literal clause-satisfaction counts (fig. 2c). Thus, although the assignments generated by the various methods contain different literals, it is not clear whether or not algorithms using sub-clause-derived assignments will outperform those using assignments generated by greedy per-literal clause-satisfaction counts, though both methods appear to outperform pseudo-random assignment generation (fig. 2c).

**Assignment Type and % Satisfied Clauses
(100 Problems, n=500, mixed satisfiable/unsatisfiable)**

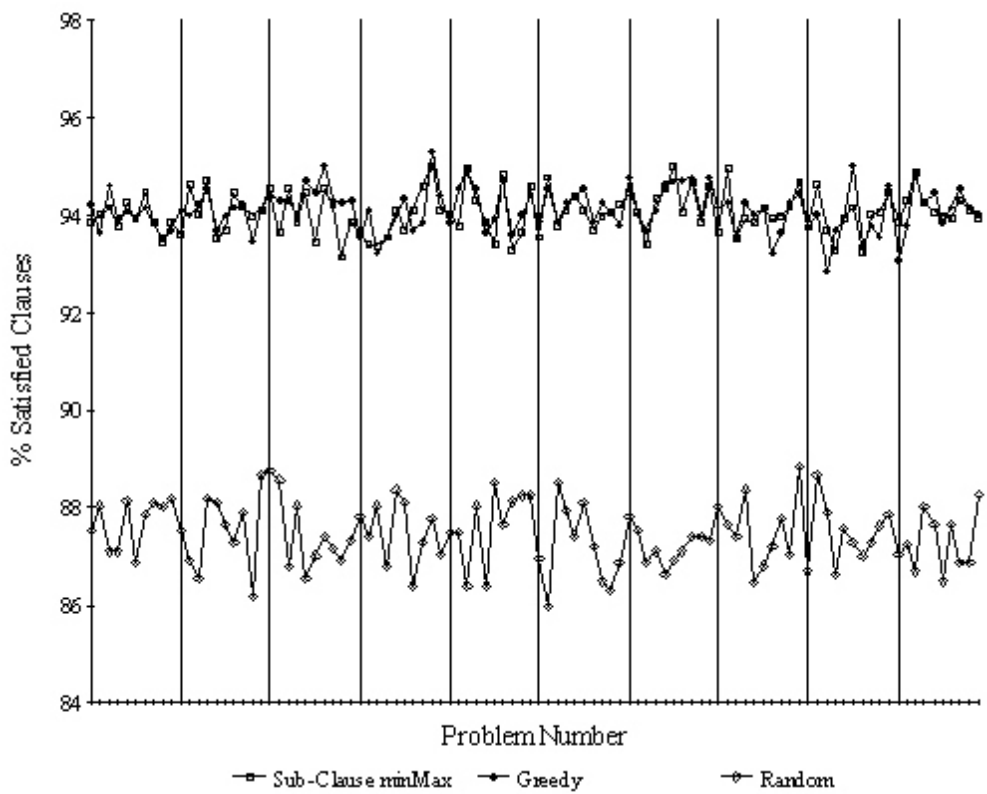

**Figure 2c.** Percent of clauses satisfied by assignments generated using either MinCreateMaxSolve (mean=94.07%), Greedy (mean=94.12 %) or Random (mean=87.45 %). (100 problems, n=500, mixed satisfiable and unsatisfiable, m=4.25. Problem set: satk3n500_218.in.)

Characteristic Inflection Point of the Unsolved Sub-Clause Curve

Sub-clauses created by a particular literal contain literals that are distributed throughout the complete set of literals.  The inflection point (the point where the sub-clauses satisfied by the assigned literals begins to exceed the sub-clauses produced by the assigned literals) of the unsolved sub-clause curve occurs, for satisfying assignments, after approximately half (n/2) of the literals are assigned (fig. 2d).

**Mean Unsolved Sub-Clauses over 54 Satisfiable Formula**

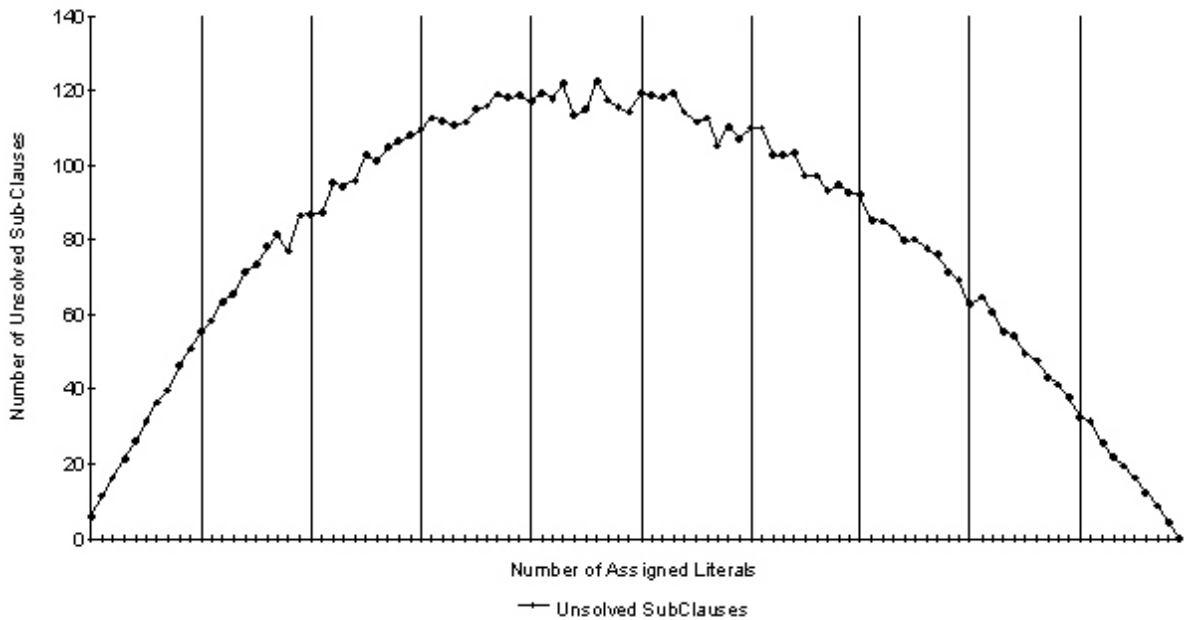

**Figure 2d.**  Mean unsolved sub-clauses at each literal in satisfying assignments for 54 satisfiable problems (n=100, m=4.25, problem sets from the k3n100Seq211 series).  The inflection point of the curve occurs at the literal where the assigned literals satisfy (consume) more sub-clauses than they create (produce), characteristically when around half the literals are assigned.

Unsatisfying Assignments and Excluded Literals

Because a satisfying assignment may not contain unsolved sub-clauses (see theorem), assignment literals that create unsolved sub-clauses must be excluded from an assignment.  Thus, because the creators of each sub-clause are known, splitting an unsatisfying assignment into an allowed and an excluded set of literals is a fairly trivial matter:

    *for (s in set of unsolved sub-clauses)*
        *excludedLiterals |= s.creatingLiteral.*

## 4. The Relation of 3-sat to 2-sat

Given an assignment $A$, each literal $a \in A$ activates the set of sub-clauses in *subclauses(a)*. Since *subclauses(a)* is a partial 2-sat problem, the conjunction of the *subclauses(a)* for all $a \in A$ is a 2-sat problem. Clearly, if a literal is used in an assignment, then its activated sub-clauses must all be satisfied (see proof). This implies that each 3-sat problem contains a set of 2-sat problems, one of which is activated by a particular assignment. For example, consider the 3-sat formula in Fig. 3, with satisfying assignment $A = \{-x_0 \ -x_1 \ x_2\}$.

**Clauses for Satisfiable 3-sat Problem $F$**
(with assignment literals $-x_0$, $-x_1$ and $x_2$ underlined):

$C_0 = \{\underline{-x_0} \ \underline{-x_1} \ -x_2\}$, $C_1 = \{\underline{-x_0} \ -x_1 \ \underline{x_2}\}$,
$C_2 = \{\underline{-x_0} \ x_1 \ -x_2\}$, $C_3 = \{\underline{-x_0} \ x_1 \ \underline{x_2}\}$,
$C_4 = \{x_0 \ x_1 \ \underline{x_2}\}$, $C_5 = \{x_0 \ \underline{-x_1} \ -x_2\}$
$C_6 = \{x_0 \ \underline{-x_1} \ \underline{x_2}\}$.

Clauses Satisfied($a$) for $a \in L$ (numbers refer to satisfied clauses):

| Satisfied for True Literals | Satisfied for False Literals |
|---|---|
| satisfied($-x_0$) = {0 1 2 3} | satisfied($x_0$) = {4 5 6 } |
| satisfied($-x_1$) = {0 1 5 6} | satisfied($x_1$) = {2 3 4} |
| satisfied($-x_2$) = {0 2 5} | satisfied($x_2$) = {1 3 4 6} |

Sub-clauses for $F$:

$S_0 = (-x_0 \lor -x_1)$, $S_1 = (-x_0 \lor x_1)$, $S_2 = (-x_0 \lor -x_2)$, $S_3 = (-x_0 \lor x_2)$,
$S_4 = (x_0 \lor -x_1)$, $S_5 = (x_0 \lor x_1)$, $S_6 = (x_0 \lor -x_2)$, $S_7 = \{x_0 \lor x_2)$,
$S_8 = (-x_1 \lor -x_2)$, $S_9 = (-x_1 \lor x_2)$, $S_{10} = (x_1 \lor -x_2)$, $S_{11} = (x_1 \lor x_2)$.

Sub-clause thresholds for $F$:

minimum_threshold = 9,
maximum_threshold = 12,
| subclauses($A$) | = 11.

Sub-clause sets, subclauses(a) for Literals $a \in L$ (numbers refer to created sub-clauses):

| True Literals | False Literals |
|---|---|
| subclauses($-x_0$) = {8 9 11} | subclauses($x_0$) = {8 9 10 11} |
| subclauses($-x_1$) = {2 3 7} | subclauses($x_1$) = {2 3 6 7} |
| subclauses($-x_2$) = {0 1 5} | subclauses($x_2$) = {0 1 4} |

**Figure 3**. Satisfiable 3-sat Problem, with associated sub-clause sets.

The assigned literals $-x_0$, $-x_1$ and $x_2$ create the following sub-clause sets:

subclauses(-x0)={8 9 11}         { S8=(-x1 v -x2), S9=(-x1 v x2), S11=(x1 v x2) }
subclauses(-x1)={2 3 7}          { S2=(-x0 v -x2), S3=(-x0 v x2), S7={x0 v x2) }
subclauses( x2)={0 1 4}          { S0=(-x0 v -x1), S1=(-x0 v x1), S4=(x0 v -x1) },

which yield the set of subclauses={0 1 2 3 4 7 8 9 11}, corresponding to the following satisfiable two-sat problem (with satisfying literals underlined):

(-x0 v -x1) ^ (-x0 v x1) ^ (-x0 v -x2) ^ (-x0 v x2) ^ (x0 v -x1) ^
(x0 v x2) ^ (-x1 v -x2) ^ (-x1 v x2) ^ (x1 v x2).

If x1 is substituted for -x1, the assigned literals resulting in the unsatisfying assignment $A=\{-x0\ x1\ x2\}$ create the following sub-clauses:

subclauses(-x0)={8 9 10}         { S8=(-x1 v -x2), S9=(-x1 v x2), S10=(x1 v -x2) }
subclauses( x1)={2 3 6 7}        { S2=(-x0 v -x2), S3=(-x0 v x2), S6=(x0 v -x2),
                                 S7={x0 v x2) }
subclauses( x2)={0 1 4}          { S0=(-x0 v -x1), S1=(-x0 v x1), S4=(x0 v -x1) },

which yield the set of subclauses={0 1 2 3 4 6 7 8 9 10}, corresponding to the following 2-sat problem (unsatisfied clauses in bold face):

(-x0 v -x1) ^ (-x0 v x1) ^ (-x0 v -x2) ^ (-x0 v x2) ^ **(x0 v -x1)** ^
**(x0 v -x2)** ^ (x0 v x2) ^ **(-x1 v -x2)** ^ (-x1 v x2) ^ (x1 v -x2).

To more formally define the relation of 3-sat to 2-sat, let *a* be an arbitrary literal, and *A* an assignment for *F*. Let *T* be the 2-sat formula derived by taking the conjunction of subclauses(a) for all $a \in A$. If *A* satisfies *F*, then *T* is a satisfiable 2-sat formula (see proof). Conversely, if *A* does not satisfy *F*, then *A* does not satisfy *T* (see corollary). More generally, any satisfying assignment of a *k-sat* formula (where *k* is an integer), generates a satisfiable *(k-1)-sat* formula. (The proof for *k-sat* follows the logic of the proof for 3-sat). It follows that a satisfying assignment for a 2-sat problem generates a satisfiable 1-sat problem which is satisfied by the satisfying assignment of the parent 2-sat problem. Thus, the 3-sat problem is inherently recursive: an assignment *A* for a 3-sat problem *F* generates a 2-sat problem *T*. The use of *A* as an assignment for *T*–a necessary condition for satisfiability of *F*–generates a 1-sat problem, which is satisfiable by *A* iff *T* is satisfiable by *A*. Likewise, *T* is satisfied by *A iff F* is satisfied by *A*.

<u>Theorem</u>: Given a satisfying assignment *A* for a 3-sat formula *F*, there is a satisfiable 2-sat formula *T* that is equal to the conjunction of the *subclauses(a)* for $a \in A$. *T is satisfied by A.*
<u>Proof</u>: Let *F* be a satisfiable 3-sat formula. Let *A* be a satisfying assignment set for *F*. For any arbitrary literal $a \in A$ there is a set of 2-sat subclauses, *subclauses(a)*. Since the clauses of *F* are joined by conjunction, and *subclauses(a)* is merely the family of subclauses given by the conjunction {satisfied(-a) ^ a}, the *subclauses(a)* are also joined by conjunction. Thus the concatenation of subclauses*(a)* for $a \in A$ is a 2-sat formula contained in *F*. Let *T* be the 2-sat formula in *F* given by $\wedge($ subclauses(a) $)$ for $a \in A$. Let *a* be an arbitrary literal in *A*. Assume that *A* satisfies *F,* but *T* is not satisfiable. If *T* is not satisfiable, then some literal $a \in A$ created an unsatisfied sub-clause $s \in T$. Since *s* is created by *a, s* is equal to some clause $c = \{s \vee -a\}$. Since $a \in A$, $-a \notin A$. Since *s* is not satisfied, $(s \wedge A) = \varnothing$, which means that $(c \wedge A) = \varnothing$, which means that *c* is not satisfied. If *c* is not satisfied, then *A* does not satisfy *F,* which contradicts the premise that *A* satisfies *F.* So, *c* must be satisfied. Since $-a \notin A$, at least one of the literals in *s* must be in A, or *c* is not satisfied. If one of the literals in *s* is in *A*, then *s* is satisfied, which means that *T* is satisfied, since *s* was the unsatisfying subclause of *T*. Thus, if *A* satisfies *F*, *A* also satisfies *T,* since otherwise there is an unsatisfied clause $c \in F$.

Corollary 1: Given an unsatisfying assignment, a subset of the generated sub-clause sets are unsatisfied by the assignment.
Proof: Let *F* be a satisfiable 3-sat formula. Let *X* be the set of variables for *F.* Let *A* be a consistent, unsatisfying assignment for *F,* containing either a true or false literal for all $x \in X$. Let $SC = \wedge (\text{subclauses}(a))$ for $a \in A$, and *C* be the set of clauses in *F.* Since *A* does not satisfy *F,* there is a clause $c \in C$ that is unsatisfied, which means that $(c \wedge A) = \varnothing$. By definition, the set of literals in *c* are true or false instantiations of three of the variables in *X*. Thus, for any literal $q \in c$, $-q \in A$. If $-q \in A$, then *subclauses(-q)* contains a sub-clause $s = (c \wedge -q)$. Since $(c \wedge A) = \varnothing$, and $s \subseteq c$, $(s \wedge A) = \varnothing$. Since $s \in SC$ and *s* is activated by some $a \in A$, it follows that in an unsatisfying assignment, a subset of the generated sub-clause sets are unsatisfied by the assignment.

Definition: Let *literals(Q)* denote the set of literals in a set of clauses *Q*.

Corollary 2: If a partial, consistent assignment *P*, which contains less than *n* literals, satisfies all of its activated sub-clauses, but does not satisfy the corresponding 3-sat formula *F*, then *F* is divisible into at least two distinct constituent 3-sat sub-formulas, *F1,* which contains the set of clauses *C1* that are satisfied by *P,* and *F2,* which contains the set of clauses *C2,* such that *literals(C2) = {L - P}*. Additionally, if $L1 = literals(C1) \cup -(literals(C1))$ and $L2 = literals(C2) \cup -(literals(C2))$, then $L1 \ne L2$.
Proof: Suppose that a 3-sat problem *F* contains the set of clauses *C,* and that *P* is a partial assignment for *F.* Suppose that *P* activates and satisfies a non-empty set of sub-clauses, but that *P* does not satisfy the corresponding 3-sat formula *F*. In other words:

$$(\forall a \in P\ (\text{subclauses}(a) = \forall a \in P(\text{subsat}(a))\ \wedge\ (\exists c \in C(P \wedge c = \varnothing)).$$

If $SC = \text{subclauses}(P)$, then *P* satisfies the set of clauses *C1,* where $C1 = \text{parent}(SC)$. Since at least one clause in *C* is unsatisfied, $C1 \subseteq C$ and $|C1| < |C|$. Let $(C2 = C - C1)$ denote the set of unsatisfied clauses of *C.* Suppose that $c = (j \vee k \vee l)$ is an arbitrary unsatisfied clause in *C*. Clearly, $c \notin C1$, since all clauses in *C1* are satisfied. Thus, $c \in C2$. Additionally, $P \wedge c = \varnothing$, since otherwise P satisfies *c*. Because $c \notin P$, ($j \notin P \wedge k \notin P \wedge l \notin P$). Similarly, ($-j \notin P \wedge -k \notin P \wedge -l \notin P$), since otherwise there would be an unsatisfied sub-clause in *C1*. Suppose, for example, that $-j \in P$. If $-j \in P$ then *P* activates a sub-clause $s = (k \vee l)$. However, s must be unsatisfied by *P,* since ($k \notin P \wedge l \notin P$), which contradicts the premise that *P* satisfies all of its activated sub-clauses. So, ($-j \notin P$), and the same is true for both -k and -l. In summary, if *F* is a 3-sat problem, *P* a partial assignment that satisfies all of its activated sub-clauses, *c* an unsatisfied clause in *F*, and *q* the set of literals in *c,* then $q \wedge P = \varnothing$ and $-q \wedge P = \varnothing$, from which it follows that *F* is divisible into at least two distinct constituent 3-sat sub-formulas, with each sub-formula satisfied (or unsatisfied) by the literals in the corresponding sub-formula clauses.

## 5. The Nested Structure of 3-Sat

If an assignment $A$ satisfies a 3-sat formula $F$, then, for an arbitrary literal $a \in A$, the sets of sub-clauses in *subclauses(a)* must be satisfied. Thus, $a$ is equivalent to ( $a$ ^ *subclauses(a)* ). Thus, for any 3-sat problem $F$, the set of literals $\{x_0, x_1, ..., x_{n-1}\}$ is equivalent to the statements in Fig. 4:

|  False Literals | True Literals |
|---|---|
| $-x_0$ ^ subclauses($-x_0$) | $x_0$ ^ subclauses($x_0$) |
| $-x_1$ ^ subclauses($-x_1$) | $x_1$ ^ subclauses($x_1$) |
| . | |
| . | |
| . | |
| $-x_{n-1}$ ^ subclauses($-x_{n-1}$) | $x_{n-1}$ ^ subclauses($x_{n-1}$) |

**Figure 4.** Literals and their associated *subclause()* sets.

Consider the literal $x_0$, equivalent to $x_0$ ^ subclauses($x_0$). Assume that subclauses($x_0$) = ($s_0$ ^ $s_1$ ^ ... ^ $s_j$). Let $L$ be the set of true and false literals and assume that $\{a_0, a_1, ..., a_i\} \in L$. Assume that $s_0 = (a_1 \vee a_2)$, and that *subclauses(a1)* = ($s_4$ ^ $s_5$ ^ $s_6$). Since $a_1 = a_1$ ^ subclauses($a_1$), and $a_2 = a_2$ ^ subclauses($a_2$), $x_0$ is equivalent to any of the following equations:

1)     $x_0 = x_0$ ^ subclauses($x_0$)
2)     $x_0 = x_0$ ^ ($s_0$ ^ $s_1$ ^ ... ^ $sj$)
3)     $x_0 = x_0$ ^ ( **($a_1$ v $a_2$)** ^ $s_1$ ^ ... ^ $sj$).

Expanding $a_1$ yields the following equation:

4)     $x_0 = x_0$ ^ ( **(($a_1$ ^ subclauses($a_1$))** ) v $a_2$) ^ $s_1$ ^ ... ^ $sj$),

and expanding *subclauses($a_1$)* yields the following equation:

5)     $x_0 = x_0$ ^ ( (($a_1$ ^ **($s_4$ ^ $s_5$ ^ $s_6$)** ) v $a_2$) ^ $s_1$ ^ ... ^ $sj$).

If $s_4 = (a_3 \vee a_4)$, then expanding $s_4$ yields the following equation:

6)     $x_0 = x_0$ ^ ( (($a_1$ ^ ( **($a_3$ v $a_4$)** ^ $s_5$ ^ $s_6$ )) v $a_2$) ^ $s_1$ ^ ... ^ $sj$),

which, since $a_3 = a_3$ ^ subclauses($a_3$), expands to the following equation:

-11-

7)     $x_0 = x_0 \wedge (\ ((a_1 \wedge (\ (\underline{(a_3 \wedge subclauses(a_3))}\ \vee a_4)\ \wedge s_5 \wedge s_6\ ))\ \vee a_2) \wedge s_1 \wedge ... \wedge sj)$.

As equations 1-7 illustrate, $x_0$ (or, by induction, any literal $a \in A$) is recursively defined in terms of itself, its set of induced sub-clauses, subclauses($x_0$), the constituent sub-clauses in subclauses($x_0$), the constituent literals of those sub-clauses, their expansions, and so on. This nesting of literals within induced sub-clause sets is one of the key features of the structure of the 3-sat problem.[1]

### 6. Literals and their Sub-clause Sets are Hypernodal Implication Graphs

Any 2-sat clause may be converted into an equivalent implication graph [8]. For example, the 2-sat clause $c=(l_1 \vee l_2)$ is equivalent to the graph containing the two implications $-l_1 \rightarrow l_2$ and $-l_2 \rightarrow l_1$. Since each sub-clause is a 2-sat clause, sub-clauses may also be converted into implication graphs (see Fig. 5)[2]. Thus, *subclauses(a)* is equivalent to $I_a$, the directed graph containing $a$ and the connected implications of the sub-clauses in *subclauses(a)*. Each (non-empty) implication graph contains instantiations of nodes that correspond to the literals in $L$. Since a node corresponds to a literal, and a literal is equivalent to an implication graph, each node reference is equivalent to an instantiation of the implication graph for the corresponding node.

Hypernodal graphs are defined as $H=\{N, G\}$, wher $N$ is the set of nodes in the graph and $G$ is a set of graphs containing the nodes. Unlike the nodes in a non-nested graph, which are terminal, the nodes in $N$ of a hypernodal graph are defined as sets of nodes and edges. In short, nodes are themselves graphs. Thus, implication graphs, whose nodes are themselves other implication graphs, are hypernodal.

Although implication graphs are hypernodal, the definition of an implication graph contains an additional component, the set of undirected cross-edges, which describes the connection of one implication graph to another. Thus, the implication graph $I_a$ of literal $a$ consists of three components: a set of nodes $N_a$, where each node $n \in N_a$ corresponds to one of the literals in subclauses($a$); a set of directed edges $E_a$, such that each edge connects a pair of nodes in $N_a$; and a set of undirected cross-edges $CE_a$, such that each cross-edge connects a node $n \in N_a$ to a corresponding node $n \in N_b$, where $b \neq a$. Thus, the implication graph for literal $a$, defined as $I_a = (N_a, E_a, Ce_a)$, describes connections between its internal nodes and the nodes of other implication graphs.

The properties of implication graphs are similar to the properties of literals and their associated sub-clause sets: the nodes in implication graphs are recursively defined (since nodes are themselves graphs) and, like the literals in sub-clause sets (see equations 1-7), recursively expandable. For example, assume that in the implication graph $I_a = (N_a, E_a, CE_a)$ there is an arbitrary node $n \in N_a$ which corresponds to a literal in $L$. Because $n \in L$, the following holds: $n$

---

[1] Although a discussion of generative grammars is beyond the scope of this paper, the growth of equations 1-7 is an example of a Lindenmeyer system[9], a type of generative grammar that may be used to model 3-sat.

[2] 3-sat sub-clauses may also be converted into finite state machines, where the antecedent of each sub-clause-derived implication induces a transition between the creator of the sub-clause and the consequent of the implication. The transition function for this finite state machine is also recursive: it is a function of its states.

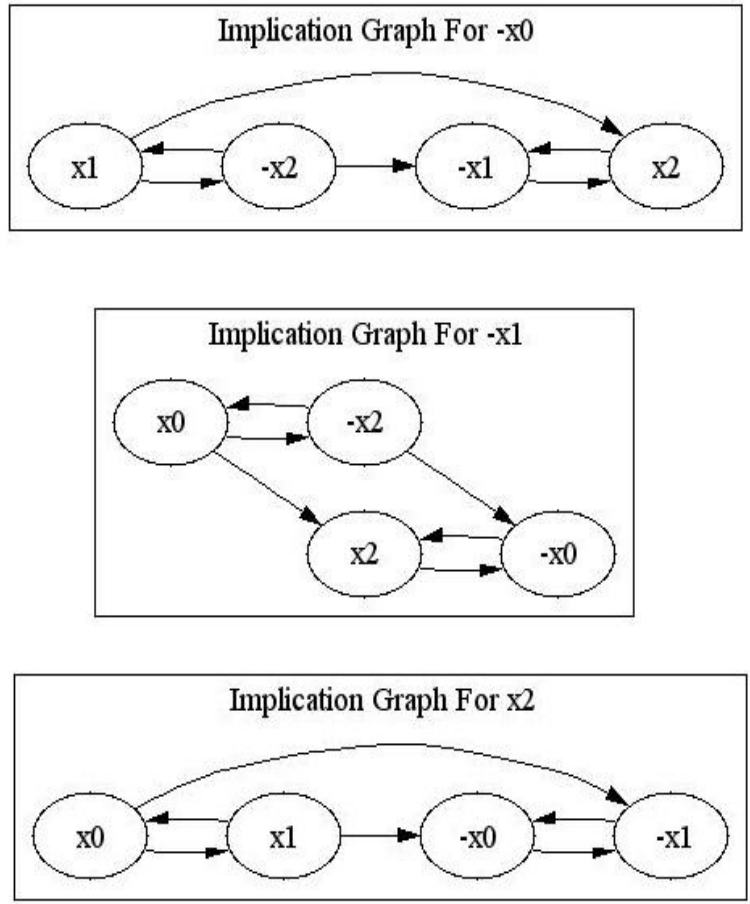

**Fig. 5:** Implication Graphs for -x0, -x1 and x2.

$= ( n \wedge \text{subclauses}(n) ) = I_n = (N_n, E_n, CE_n)$. Thus, $n$ is both the graph $I_n = (N_n, E_n, CE_n)$ and a node in the graph $I_a = (N_a, E_a, Ce_a)$. In other words, implication graphs contain nodes which are themselves implication graphs. This dual nature of nodes as both nodes and graphs aptly reflects the nested structure of 3-sat. In summary, the hypernodal (nested graph) model may be used to represent 3-sat problems.

### 6. The Hypernodal Model of 3-Sat

Since any 3-sat problem in n variables contains $2n$ literals, and each literal has a corresponding hypernodal implication graph, there are $2n$ hypernodal implication graphs per 3-sat problem. Thus, a 3-sat formula $F$ may be represented by the set of hypernodal implication graphs, $HG=\{I_{-x0}, I_{-x1}, ..., I_{-xn-1}, I_{x0}, I_{x1}, ..., I_{xn-1}\}$, corresponding to the literals in $F$.

If the implication graphs in *HG* are grouped to correspond to true and false literals, the hypernodal representation of a 3-sat problem *F* looks like two stacks of directed graphs, one stack for the set of true implication graphs $\{I_{x0}, I_{x1}, ..., I_{xn-1}\}$ and another for the set of false literal graphs $\{I_{-x0}, I_{-x1}, ..., I_{-xn-1}\}$, with undirected connections between the graphs representing cross-edges (see Fig. 6).

**Set of Hypernodal Graphs *HG*, Arranged by Literal**

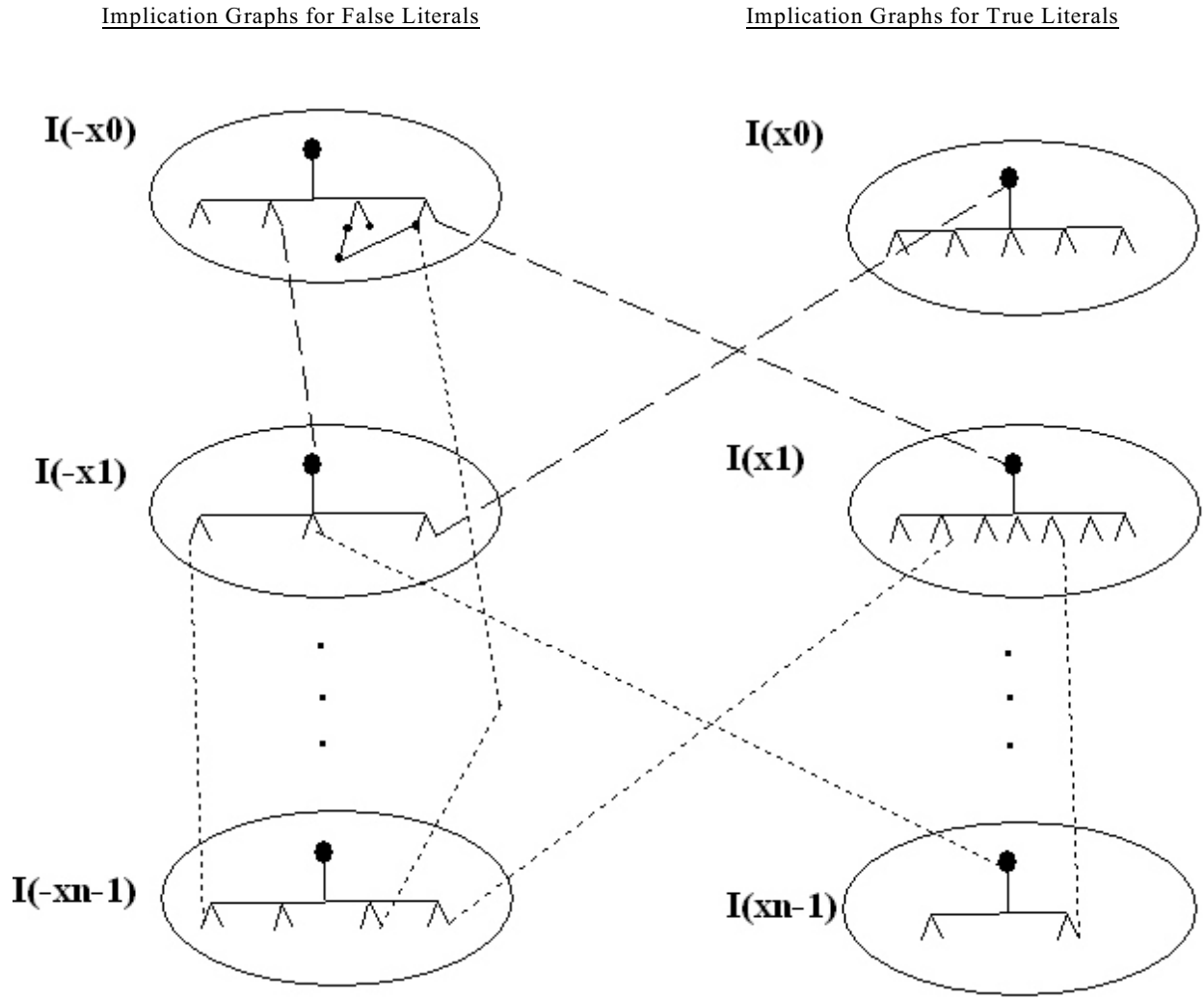

**Figure 6.** The set of hypernodal implication graphs for a 3-sat problem. Each ellipse contains an implication graph $I(x_i)$ for the corresponding literal. Blackened circles are the upper-level nodes that correspond to literals. The other nodes are references to literals and thus contain other implication graphs. A dashed line between an upper level node and a leaf node indicates that the leaf node contains the upper level graph. A dotted line between two leaf nodes is an undirected cross-edge. Cross-edges activate endpoint nodes when the implication graphs containing the endpoint are activated (used in an assignment).

## 7. The Mechanism of Success and Failure in 3-Sat Assignments

The use of two literals *a* and *b* in an assignment induces the sets of subclauses corresponding to *a* and *b*. Since *subclauses(a)* is equivalent to $I_a$ and *subclauses(b)* is equivalent to $I_b$, the subclauses of the set $\{a \cup b\}$, *subclauses(a $\cup$ b)*, is equivalent to $I_a \cup I_b$. So, given an assignment *A,* for each literal implication graph $I_a$ corresponding to each literal $a \in A$, the nodes $n \in \{N_a \wedge A\}$ are active, so the successors of *n* must also be included in the assignment. Similarly, if a cross-edge *CE* from $I_a$ reaches another active literal implication graph $I_b$, the reached node (and its successors) in $I_b$ must be included in the assignment. Likewise, the negation of any of the literals in *A* may not be reached by any of the nodes in the active implication graphs, or a contradiction results. In short, the effect of using a set of literals in an assignment is to combine the implication graphs of the set of literals.

Unsuccessful assignments create combinations of implication graphs whose merged nodes reach contradictions (i.e. the paths contain strongly-connected inconsistencies), whereas the active paths induced by successful assignments are consistent. This implies that the upper-bound on the cost of solving the 3-sat problem is proportional to the cost of performing any of the three following operations on the set of hypernodal graphs that represent a 3-sat formula: 1) calculating the transitive closure, 2) calculating the strongly-connected components, or 3) recognizing and cutting contradictory paths from the graph.

## 8. Conclusion

Decomposing 3-sat problems into sub-clause-derived hypernodal graphs elucidates the recursive structure of 3-sat and allows one to easily visualize otherwise complex 3-sat problems. Sub-clause-derived hypernodal graphs aptly describe the mechanism by which particular assignments either succeed or fail.

The performance of existing satisfiability algorithms may be improved by the use of various types of sub-clause-derived assignments as a starting point for satisfiability searches. Sub-clause thresholds and the comparison of produced and consumed sub-clauses may prove to be valuable measures that algorithm designers may use to design new satisfiability algorithms. Additionally, using sub-clauses to extract excluded literals from unsatisfiable assignments may allow algorithm designers to design algorithms that better focus the search for satisfiable assignments, thereby improving algorithm performance.

Further investigation into the complexity of calculating the transitive closure or finding the strongly connected components of hypernodal graphs will cast light on the upper-bounds of the complexity of the 3-sat problem. In summary, further investigation of sub-clauses, hypernodal graphs and 3-sat may open new avenues for 3-sat research and help algorithm designers develop more optimal 3-sat algorithms.

## References

1. A. Poulovassilis and M. Levene, A Nested Graph Model for the Representation and Manipulation of Complex Objects. *ACM Transactions on Information Systems (TOIS)*, 12 (1), 35-68 (1994).
2. T.H. Cormen, C.E. Leiserson, R.L. Rivest, *Introduction to Algorithms*. The MIT Press. Cambridge, Massachusetts, 1997.
3. J. Gu, P.W. Purdom, J. Franco and B. W. Wah, Algorithms for the Satisfiability (SAT) Problem: A Survey (Preliminary Version, 2000), *DIMACS Series in Discrete Mathematics and Computer Science*.
4. B. Aspvall, M. F. Plass and R. E. Tarjan, A Linear-Time Algorithm for Testing the Truth of Certain Quantified Boolean Formulas. *Information Processing Letters* 8 (3), 121-123 (1978).
5. R. Monasson, R. Zecchina, S. Kirkpatrick, B. Selman and L. Troyansky. Determining Computational Complexity from Characteristic 'Phase Transitions.' *Nature*, 400, 133-137 (1999).
6. B.Selman, D.G. Mitchell, H.J. Levesque, Generating hard satisfiability problems. *Artificial Intelligence*, 81, 17-29 (1996).
7. I.P. Gent and T. Walsh, The Search for Satisfaction, *Internal Report, Dept. of Computer Science*, University of Strathclyde, 1999.
8. Van Gelder, A., Tsuji, Y.K.: Satisfiability Testing with More Reasoning and Less Guessing. In Johnson, D., Trick, M., eds.: *Cliques, Coloring and Satisfiability. Volume 26 of DIMACS Series in Discrete Mathematics and Theoretical Computer Science*. American Mathematical Society (1996) 559--586.
9. Dewdney, A. K., *The Turing Omnibus: 61 Excursions in Computer Science*, Computer Science Press. Rockville, Maryland, 1989.